\documentclass[10pt,twocolumn,letterpaper]{article}

\usepackage[pagenumbers]{cvpr} 

\usepackage{epigraph}
\usepackage{graphicx}
\usepackage{amsmath}
\usepackage{amssymb}
\usepackage{booktabs}
\usepackage{tabularx}
\usepackage{makecell}

\usepackage[pagebackref,breaklinks,colorlinks]{hyperref}

\usepackage[capitalize]{cleveref}
\crefname{section}{Sec.}{Secs.}
\Crefname{section}{Section}{Sections}
\Crefname{table}{Table}{Tables}
\crefname{table}{Tab.}{Tabs.}

\def\cvprPaperID{*****} 
\def\confName{CVPR}
\def\confYear{2022}

\newif\ifshowauthorcomments

\showauthorcommentstrue

\newcommand{\manipulation}{{manipulation}\xspace}

\newcommand{\stkout}[1]{\ifmmode\text{\sout{\ensuremath{#1}}}\else\sout{#1}\fi}

\newif\ifseechanges

\ifseechanges

\else

\fi

\begin{document}

\title{Zero-shot CAD Program Re-Parameterization for Interactive Manipulation}

\author{Milin Kodnongbua*\\
University of Washington\\
{\tt\small milink@cs.washington.edu}
\and
Benjamin T. Jones*\\
University of Washington\\
{\tt\small benjones@cs.washington.edu}
\and
Maaz Bin Safeer Ahmad\\
Adobe Research\\
{\tt\small mahmad@adobe.com}
\and
Vladimir G. Kim\\
Adobe Research\\
{\tt\small vokim@adobe.com}
\and
Adriana Schulz\\
University of Washington\\
{\tt\small adriana@cs.washington.edu}
}
\twocolumn[{%
\renewcommand\twocolumn[1][]{#1}%
\maketitle
\begin{center}
    \centering
    \captionsetup{type=figure}
    \includegraphics[width=\textwidth]{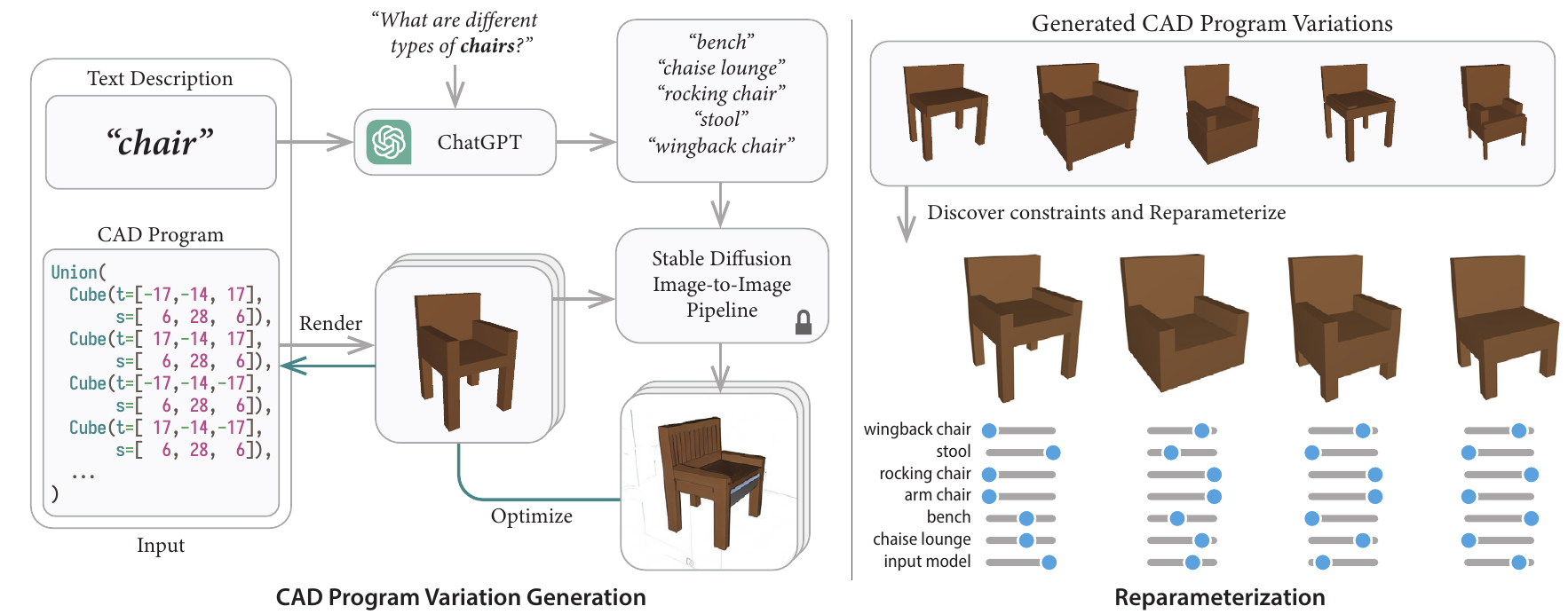} 
    \caption{Our method takes a parametric 3D model and a corresponding text description as input and generates a reparameterized version of the model as output. It uses a language model to generate text prompts describing variations of the input model and uses a text-to-image model to optimize the CAD parameters towards these prompts. The resulting variations are used by the constraint discovery system to identify common constraints across all designs.}
    \label{fig:teaser}
\end{center}%
}]

\begin{abstract}
  Parametric CAD models encode entire families of shapes that should, in principle, be easy for designers to explore. However, in practice, parametric CAD models can be difficult to manipulate due to implicit semantic constraints among parameter values. Finding and enforcing these semantic constraints solely from geometry or programmatic shape representations is not possible because these constraints ultimately reflect design intent. They are informed by the designer's experience and semantics in the real world. To address this challenge, we introduce a zero-shot pipeline that leverages pre-trained large language and image model to infer meaningful space of variations for a shape. We then re-parameterize a new constrained parametric CAD program that captures these variations, enabling effortless exploration of the design space along meaningful design axes.
\end{abstract}

\section{Introduction}

\epigraph{The goal is to create a system that would be flexible enough to encourage the engineer to easily consider a variety of designs. And the cost of making design changes ought to be as close to zero as possible.}{\textit{Samuel Geisberg, PTC Founder, 1988}}

The impact of Parametric CAD on engineering design cannot be overstated. Almost every manufactured object that exists today started its life in a parametric CAD tool. Yet, decades past the CAD revolution, the envisioned goals of effortless design variability and \manipulation remain unrealized. The foundational vision of parametric CAD is to enable \manipulation through parameter tweaking within a sequence of parametric constructive operations. For instance, if we model the back of a chair as an extrude of a base 2D rectangle, we can manipulate the height of the chair by varying the extrude distance. In practice, modifying CAD parameters directly can be challenging due to a lack of user understanding regarding which parameters to modify to achieve the desired variation. In addition, the absence of constraints can lead to undesirable shape changes that violate user intent.

To address the gap in CAD \manipulation, this paper aims to automatically construct a reparametrization of CAD models, introducing what we refer to as \emph{\manipulation} parameters. The parameters generated by sequences of constructive CAD operations will be referred to as \emph{constructive} parameters to differentiate them from the proposed abstraction. We observe that CAD programs are typically overparameterized, as multiple constructive operations may be required to create structurally related geometries. However, constraints across constructive parameters are frequently absent or underspecified during the CAD modeling sequence, and feasible ranges for constructive parameters are not exposed. Consequently, modifying a single CAD constructive parameter can lead to shapes that lack the essential structure, such as a chair with its legs disconnected from its base. Thus, to achieve meaningful design variations, users often must simultaneously and consistently modify multiple constructive parameters~\cite{Thefaile61:online}. Essentially, the space of shape variations defined by constructive parameters predominantly consists of irrelevant outcomes, rendering the extraction of meaningful design variations from this space an arduous and time-consuming process. This work explores the possibility of automatically identifying a constrained subspace within a CAD program that reflects meaningful shape variations. We frame this as a reparametrization problem from \emph{constructive} to \emph{\manipulation} parameters.

On one hand, we anticipate that program analysis will shed light on the constraints to consider when constructing this subspace. On the other hand, this problem is inherently ambiguous: meaningful design variations are ultimately derived from \textit{what} the designer is trying to model, rather than \textit{how} they are modeling it. Any synthetic analysis would inevitably fail to infer semantic meaning. Therefore, our key insight is to first develop an understanding of how we may want to manipulate the shape and subsequently conducting an analysis based on that understanding to derive a constrained shape space. We note that establishing this understanding is now possible because of novel pre-trained large foundational models \cite{rombach2022highresolution} that have learned the space of reasonable shapes. Building upon this insight, our neurosymbolic approach combines AI-driven induction for discovering shape variations with symbolic-driven deductive reasoning for identifying shape constraints. Most notably, our approach operates in a `zero-shot' manner, eliminating the reliance on large categorical shape datasets. As such, our method is applicable beyond the sparse shape categories covered by existing datasets.

Our novel system, illustrated in Figure~\ref{fig:teaser}, takes a parametric model expressed in a simplified CAD language, along with a concise text description of the model. As output, it generates a re-parameterized version of the input CAD model, accompanied by an intuitive slider-based interface. The slider interface allows users to vary the \manipulation parameters introduced in the revised CAD program, empowering them to easily explore meaningful design variations. 

The re-parameterization process begins by using a language model to generate text prompts describing the variations of the model. Next, we apply our novel method to automatically adjust the parameters of the input model, aligning it with each text prompt by comparing the rendered images of the model with images from a pre-trained stable diffusion text-to-image model. The design variations generated by matching the input model to various text prompts are then fed into our constraint discovery system. This identifies geometric constraints that are common across all variations, accounting for noise. The discovered constraints are used to construct a subspace of the original CAD parameter space, automatically imposing semantically meaningful constraints. Finally, we project the generated variations into this subspace and use them as the basis for a new parameterization of the constrained space along semantic lines.

We demonstrate the efficacy of our approach in generating variations for five distinct models of varying complexity. Furthermore, we conduct a comparative analysis between our neurosymbolic approach and purely symbolic or purely neural-driven methods to underscore the advantages inherent in our approach; the former creates uninteresting variations and the latter can produce incoherent geometry, but our approach produces interesting variations while adhering to semantically meaningful constraints.

\section{Related Work}

Our work builds upon a rich body of work on CAD \manipulation as well as structure-preserving shape \manipulation. This includes algorithms that operate on input shape collections of a specific class, as well as methods for manipulating individual models. Furthermore, our work builds on text-driven shape generation algorithms. 

\subsection{Parametric CAD Manipulation}
Parametric CAD systems represent designs as programs that expose constructive parameters. Manipulating CAD models solely by adjusting these parameters can be challenging due to the need for coordinated changes across multiple parameters to achieve specific design goals~\cite{failedpromise}. This has encouraged efforts that diverge from the parametric modeling paradigm, proposing interfaces that allow \emph{direct \manipulation} instead (e.g. SpaceClaim, KeyCreator, and Rhino). Nevertheless, most CAD systems prioritize preserving program information as it enables the preservation and control of global structures (e.g., Solidworks, Onshape, Catia, Creo and NX).

Recent efforts have focused on exploring hybrid techniques that aim to bridge the gap between the parametric programming paradigm and direct \manipulation approaches. This includes commercial systems like Siemens' Synchronous Technology and IronCAD, which aim to facilitate direct manipulation by utilizing complex algorithms to maintain synchronization with the program representation. Efforts within the computer graphics community have made advancements in enabling program updates based on user interactions~\cite{cascaval2022bidirectional,michel2021dag}, optimizing program parameters to align with user \manipulation. The fundamental challenge with these approaches is that they rely on hand-crafted heuristics to resolve the inherent ambiguities in the system. There are often multiple viable constraints that can be imposed over the program parameters to achieve the edits that adhere to users' \manipulation. To eliminate the need for hand-crafted heuristics, we propose leveraging semantic understanding extracted from large pre-trained foundational models. By utilizing these models, we can uncover a meaningful set of potential variations for a CAD model and derive constraints directly from those examples.

\subsection{Enabling Manipulation from Large Shape Collections}
Considerable research has been conducted to explore methods for understanding the meaningful space of variations within categorical shape collections. These approaches either create an abstraction for a collection of shapes belonging to a specific category or enable manipulation of an image based on a collection of models within that category.

Early approaches combine statistical models, with label-driven shape decomposition~\cite{chaudhuri2013attribit,fish2014meta,ovsjanikov2011exploration}. Additionally, labels have been used to learn semantic abstractions from shape collections~\cite{yumer2015semantic}.

More recently, neural networks have been used to learn embeddings that enable design exploration. These approaches do not require segmented labels, but the learned embeddings are not easy for a user to explore~\cite{chen2019learning,zheng2022sdf}. To enable user control, several approaches have explored integration of learning with structural, compositional, and symbolic abstractions. This includes efforts focused on learning abstractions~\cite{jones2022neurally, tulsiani2017learning}, fitting shapes to categorical abstractions~\cite{pearl2022geocode,wei2020learning}, enabling structure manipulation through handles or mixing and matching~\cite{mo2019structurenet,hao2020dualsdf,yin2020coalesce,jiang2020shapeflow,hertz2022spaghetti,liu2021deepmetahandles}, and text-driven variations~\cite{achlioptas2022changeit3d}.

Closest to our approach are methods that infer higher-level abstractions that preserve program structure~\cite{jones2021shapemod,jones2023shapecoder}. Library learning techniques, based on machine learning~\cite{dreamcoder} or anti-unification~\cite{babble}, can extract common structure from a corpus of CAD programs into reusable functions that expose more semantically meaningful parameters.

The fundamental limitation of these approaches is that they require a large dataset of models belonging to a specific class, from which meaningful space of variations can be inferred. Rather than being confined to specific classes of objects that are covered by existing datasets, we leverage the much broader understanding embedded in foundational models to infer meaningful variations from a single input model.

\subsection{Structured Manipulation from Single Input Shapes}
Past research has also devised methods for structure-preserving shape manipulation when only a single input shape is available. Earlier methods used hand-crafted heuristics with numerical optimization to enable shape deformation (see~\cite{mitra2014structure,sorkine2009interactive} for a more complete overview). While some geometric and physics-inspired heuristics are well suited for organic shapes~\cite{igarashi2005rigid,sorkine2004laplacian}, heuristics based on geometric-semantic constraints such as symmetry, coplanarity, and replicable patterns have been shown to work well on man-made shapes~\cite{bokeloh2012algebraic,gal2009iwires}. These heuristics have been used in two types of editing systems: 1) \emph{variational methods}, where  optimization is used to compute a deformation that adheres to the user manipulation~\cite{sumner2007embedded}; and 2) \emph{direct methods} where the computation is done in advance to generate a set of exposed controls~\cite{jacobson2011bounded}. Such controls can be in the form of parameter sliders, cages, skeletons, or compositions thereof. Essentially, direct methods generate a type of reparametrization, as we describe in this work. 

More recent approaches have looked at applying learning approaches for manipulating shapes. However, while approaches that assume categorical data sets can use learning to replace heuristics~\cite{sung2020deformsyncnet}, as discussed above, efforts that take a single shape as input are more restricted. Most successful efforts focus on a constrained task of matching the input model to a target shape or image~\cite{wang20193dn,wang2018pixel2mesh}. Some efforts have been made on learning abstractions that are category independent, such as learning to fit cages~\cite{yifan2020neural} and inferring 3D shape programs from a target image or 3D geometry ~\cite{Du:2018:IverseCSG,nandi2018functional,Yu_2022_CVPR,jones2020shapeassembly} (see ~\cite{ritchie2023neurosymbolic} for a complete overview). Similarly, domain-specific compilers have been developed that strive to reduce the number of parameters in the model by re-writing CAD programs concisely using looping constructs~\cite{nandi2020synthesizing}. Such methods still focus on lower-level abstractions, essentially producing the types of programs we take as input.

\subsection{Text-conditioned 3D generation}
Several studies have explored text-conditioned 3D generative models trained directly on text-3D pairs. Most of these approaches rely on learning 3D latent representations and establishing associations between text and 3D embeddings~\cite{chen2018text2shape, liu2022implicit, sanghi2022clipforge, zeng2022lion, mittal2023autosdf, fu2023shapecrafter, sanghi2023clipsculptor}. However, scaling these methods to accommodate diverse text prompts is difficult due to the lack of large-scale 3D datasets.

A growing body of research focuses on text-conditioned 3D generation, leveraging pretrained text-to-image models like CLIP \cite{radford2021learning}, as well as diffusion-based models \cite{nichol2022glide, rombach2022highresolution, saharia2022photorealistic}. Differentiable rendering techniques are also used to optimize 3D representations such as meshes \cite{Mohammad_Khalid_2022} and NeRFs \cite{jain2022zeroshot, poole2022dreamfusion,lin2023magic3d}. However, these methods tend to struggle to generate coherent 3D objects due to lack of strong 3D priors. Recent approaches have attempted to solve this problem by generating a synthetic dataset of image-3D pairs. These methods use learning to generate the initial coarse 3D objects from images, which then serve as a starting point for further refinement through fine-grained 3D shape optimization~\cite{xu2023dream3d, nichol2022pointe, seo2023let}.

To the best of our knowledge, our work is the first to tackle text-conditioned 3D synthesis within the domain of CAD programs. While we also leverages differentiable rendering and diffusion-based models, our focus lies on the problem of distilling the inherent constraints on CAD parameters from a large pretrained model.

\section{Methods}
Given a well-formed CAD model represented as a simplified constructive solid geometry (CSG) and a categorical description of the model (``chair''), our system synthesizes a new CAD program with fewer parameters capable of reproducing the input space and expressing meaningful semantic variations. We first use a large language model to generate text description of variations of the given object (``bench'', ``stool'', etc.). We then optimize the parameters of the input CAD program to fit these variation prompts using diffusion-generated images as a guidance to a differentiable renderer. From these instances of model variations, we infer constraints and reparameterize to a CAD program that exposes meaningful \manipulation parameters to the users.

\subsection{Simplified CAD Language}
Our simplified CAD language is built upon Constructive Solid Geometry (CSG), which forms the basis of popular software like OpenSCAD, and is a supported mode of operation in most commercial CAD systems. While modern CAD systems predominantly employ B-rep history-based languages, the essential boolean operations of CSG persist. To simplify the implementation of a differentiable renderer, we have opted to include only union operators in our language, excluding intersections and subtractions. Despite this restriction, our language still allows for a wide range of CAD designs, showcasing the capabilities of our method.  Specifically, we have implemented operators for three primitives (cubes, cylinders, cylinders with changing top radius), as two transformation operators that can be applied to any primitives (translation and scale).

The constructive parameters of the language include transformation parameters and primitive-dependent parameters (e.g., the top radius of a cylinder).  It is worth mentioning that numerous CAD systems allow users to expose high-level variables and define constructive parameters as functions of these variables. However, in our approach, we do not assume the presence of such variables in the input. Instead, we focus on learning high-level abstractions directly from the constructive parameters themselves, highlighting the effectiveness of our method in uncovering meaningful constraints.

\subsection{Variation Prompts Generation}

We leverage a pre-trained LLM (ChatGPT) to generate text prompts that describe variations of the given model by using the following formulaic query: ``What are different types of [object]?''. This generally outputs an itemized list of prompts which we can extract. A user can then select a subset of these prompts and add additional prompts that they care about.

\subsection{Text-Conditioned Variation Generation}

In this section, our goal is to generate variations of the initial CAD model to serve as examples to discover constraints and re-parameterize the CAD program in a semantically meaningful way. For each text prompt, we follow a general framework to generate 3D shapes from a pre-trained text-image model where it uses a differentiable pipeline to bridge a 3D representation to an image.

A fundamental challenge for our domain is that the space of possible variation of a CAD program is highly constrained compared to meshes or neural radiance fields (NeRFs). Therefore, to optimize a CAD program's parameters from text-driven image guidance, we constantly need to project the target variations back to the feasible space. We notice that losses used in prior work such as ClipMesh \cite{Mohammad_Khalid_2022} and DreamFusion \cite{poole2022dreamfusion} create artifacts in our application, essentially disconnecting CAD primitives (see discussion in Section~\ref{sec:ablation}). We attribute these errors to the challenges of finding the proper projection back to the CAD domain. To overcome this challenge, we propose to use an image-space loss because the transformations between CAD parameters, meshes, and images are more straightforward, facilitating the projection process. We consistently observe that this approach significantly improves the overall results.

\begin{figure}
    \centering
    \includegraphics{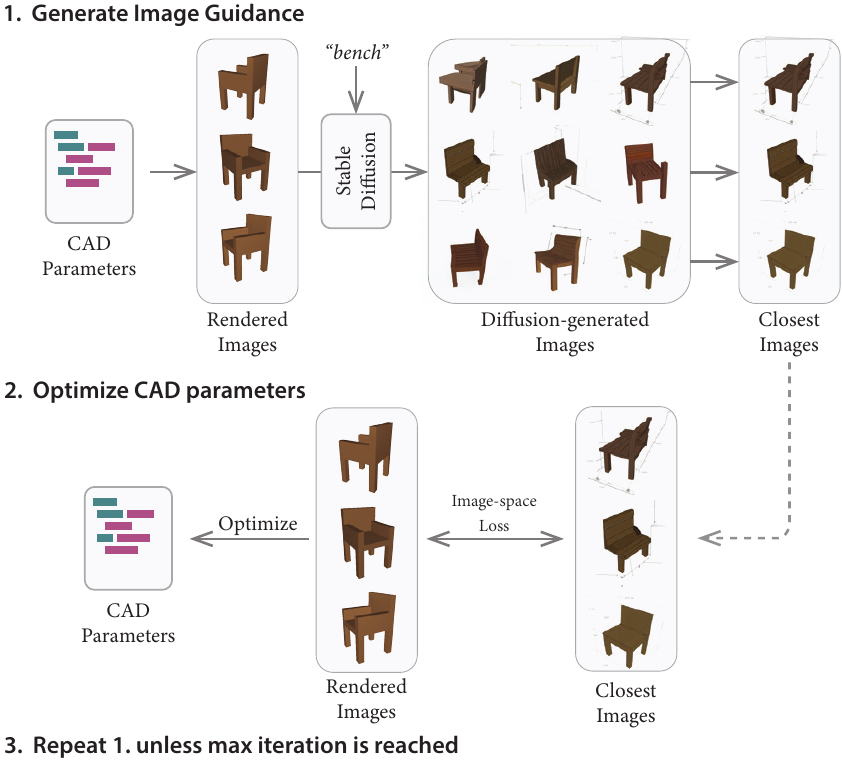}
    \caption{Text-conditioned design variation. We iteratively generates images using stable diffusion and refine our model to match the sampled image.}
    \label{fig:generation-overview}
\end{figure}

Our generation algorithm (Figure~\ref{fig:generation-overview}) iteratively generates images using stable diffusion~\cite{rombach2022highresolution} conditioned on the input prompt and the current rendering of the CAD model. After each sample, it projects back to the CAD parameters by fitting our input model to the sampled image using gradient descent on pixel loss from a differentiable renderer~\cite{Laine2020diffrast}. 

In generating the image guidance, we randomly sample 5 camera angles from which we render each image and run the image-to-image diffusion. Similarly to DreamFusion \cite{poole2022dreamfusion}, we also append a viewpoint description to the prompt, e.g., ``chair, (front|side|rear) view'' as a proxy to condition camera angles. 

In the gradient descent steps, we use an SGD optimizer with a learning rate of 0.05 for 30 iterations where we minimize the following loss:
\begin{align*}
    \mathcal{L}(\mathbf{x}) := \sum_a \lVert \text{sharpen}(R_a(\mathbf{x})) - \text{sharpen}(I_a) \rVert^2 + \lambda \lVert \mathbf{x} - \mathbf{x}_0 \rVert^2
\end{align*}
where $\mathbf{x}$ is the CAD parameters, $a$ is a camera angle, $R_a(\cdot)$ is the renderer, $I_a$ is the image from the diffusion process, $\text{sharpen}(I) := I - 0.2 \cdot \text{blur}(I)$, and $\lambda = 0.001$ is the regularization term towards the original parameters $\mathbf{x}_0$. We repeat the process of diffusion and gradient descent 400 times.

A final fundamental challenge we encounter is that diffusion models can produce images with diverse styles and conflicting geometries across different views.
This can lead to degenerate results, such as a table with thin metal legs disappearing because when the legs appear at different positions in the generated images, the optimizer does not know which legs to follow and so makes it invisible instead (also see Section~\ref{sec:ablation}).
To overcome this, we employ a selection process where we run diffusion multiple times to generate a set of images and then choose the one that aligns best with the initial rendered image to help improve geometric consistency across views. This selection is performed at each step to provide effective image guidance. We have observed that this approach consistently produces favorable outcomes.

\subsection{Constraint Discovery}
Our generative model produces a collection of examples in the design space we wish to explore. We want to use these examples to provide structure and constraints on that design space to aid in exploration by discovering constraints on the CAD model parameters that are common to the discovered variations. While there are many existing methods for constraint discovery~\cite{constraintlearning,constraintsfromexamples,invsygus}, most of these methods require noise-free examples, which we do not have, except for on our input model. For this reason, we propose algorithms for selecting among possible constraints under the presence of noise.

\subsubsection{Discovering Geometric-Semantic Constraints}
\label{{sec:applyconstraints}}

Because we have only a single model with clean geometry, our initial input, we propose to discover common geometric relationships -- coplanarity, coaxiality, keypoint coincidence, and dimensionaly equality -- present within the initial model, which we represent as conjunctions of linear constraints. The subspace these induce is too specific to the input model, so we would like to find a subset of these constraints that is common (in approximation to handle noise) to all discovered variations.

When dealing with constraints, it is crucial to consider their potential interactions. Ideally, we would rank all possible combinations of constraints; however, this combinatorial problem is computationally intractable, so we propose a greedy strategy instead. Using a distortion metric (described below), we score every individual constraint and initialize our constraint set with the best one. We then iteratively add to this set by scoring each of the remaining unused constraints unioned with the constraints already chosen. The result of this process is an ordered set of constraints from first to last picked. Plotting the distortion over number of constraints, we observe that there is usually a point where the distortion increases drastically and we have observed that this jump in the graph typically correlates to visual artifacts in the shape. We use this to compute the cutoff of which constraints to impose. We can find the jump in the graph using a linear change point detection algorithm ~\cite{killick2012optimal} on the derivative of this curve (computed as a central difference). 

To score a candidate constraint set, we would ideally measure the distance in pixel space to the diffusion images from the final stage of generation. This requires solving an optimization problem to find the minimizer of that distance over the CAD parameter space, which is too slow to be tractable with the greedy strategy described above. We also observe that pixel space differences can be unreliable for very small variations arising from small constraint sets because capturing them in an image is highly dependent on viewing angle. To overcome both of these issues, we propose to instead use a volumetric score, intersection over union (IoU), when initially sorting constraints, and only use pixel space loss when computing the final cutoff. Because we do not have a way to differentiably compute shape booleans, and to gain computational efficiency by avoiding gradient descent entirely, we developed a linear cuboid approximation to IoU which, in conjunction with our linear constraint sets, allows us to minimize the IoU with a single linear least squares step (see supplemental material for details). In cases where the IoU simplification does not generalize (cones with variable half-angle) we fall back to image loss for the full pipeline and avoid the additional check for constraint combinations due to its computational complexity.

\subsubsection{Discovering Discrete Variations}

Not all examples of the same kind of object have the same parts; for example some cameras have flash bulbs while some do not, and chairs can be armless. We discover these discrete parameters by looking for primitives effectively missing from variations. We iteratively remove one primitive from each variation and compute the pixel loss described above. If this loss is below a threshold ($10^{-4}$ in our experiments), we mark that primitive as optional for that variation. We then group parts that are always optional together across variations (e.g. chair legs are added or removed as a set) and include these as binary variables on top of our continuous reparameterization.

\subsection{Re-Parameterization}
Our generation and constraint discovery algorithms find a set of linear constraints which restrict the construction parameters to semantically appropriate values, as well as a set of semantically labeled variations that obey these constraints. The constraints give us a lower dimensional subspace that maintains object coherence. Using Gauss-Jordan elimination we can construct a basis for this subspace that retains parameter identity for free variables under the constraints. We construct a second parameterization of this space centered at the initial model's parameters within the subspace that allows these shapes to be mixed and interpolated.

\begin{figure}
    \centering
    \includegraphics[width=\linewidth]{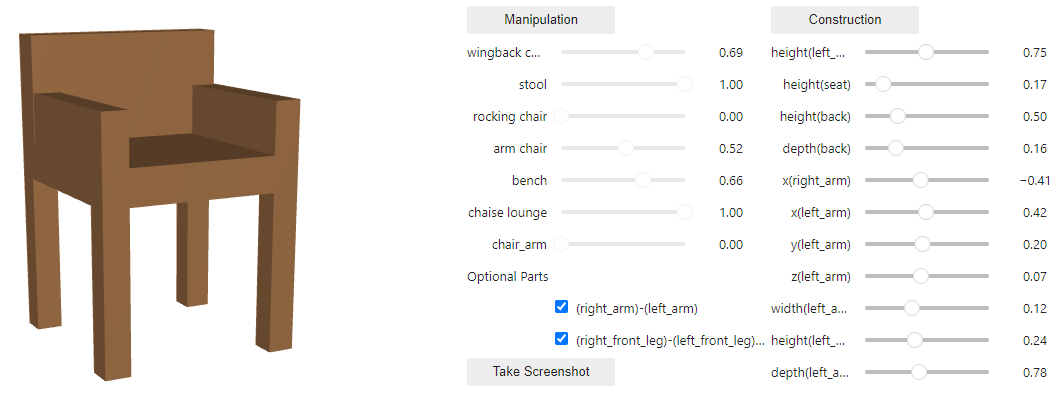}
    \caption{Our user interface. Users are presented with two complimentary views of the constrained reparameterization space. The sliders on the left interpolate between discovered variations as a weighted average, while those on the right control the free variables after reparameterization. Checkboxes allow toggling of discrete sets of removable parts.}
    \label{fig:ui}
\end{figure}

Our user interface (Figure~\ref{fig:ui}) allows for exploration in both \textit{constructive} parameters and \textit{\manipulation} parameters, but because we only found linear equality constraints, we do not have bounds on this space. We again use examples as lower bounds on the feasible region by taking a normalized sum of manipulation parameters, restricting semantic variations to be interpolations between variations, and optionally restricting constructive parameter exploration to the extreme values of the discovered variations. In this bounded subspace, the user can freely and safely explore. Additionally, we add the discrete optional degrees of freedom found during constraint discovery.

\section{Results}\label{sec:results}
In this section, we evaluate the performance of our method in two key aspects. First, we assess its ability to generate compelling and valid design variations from the input 3D model. Second, we examine its ability to infer semantic and geometric constraints that define the family of shapes described by the input model and textual prompts.

\begin{figure*}[ht!]
    \centering
    \includegraphics[width=\linewidth]{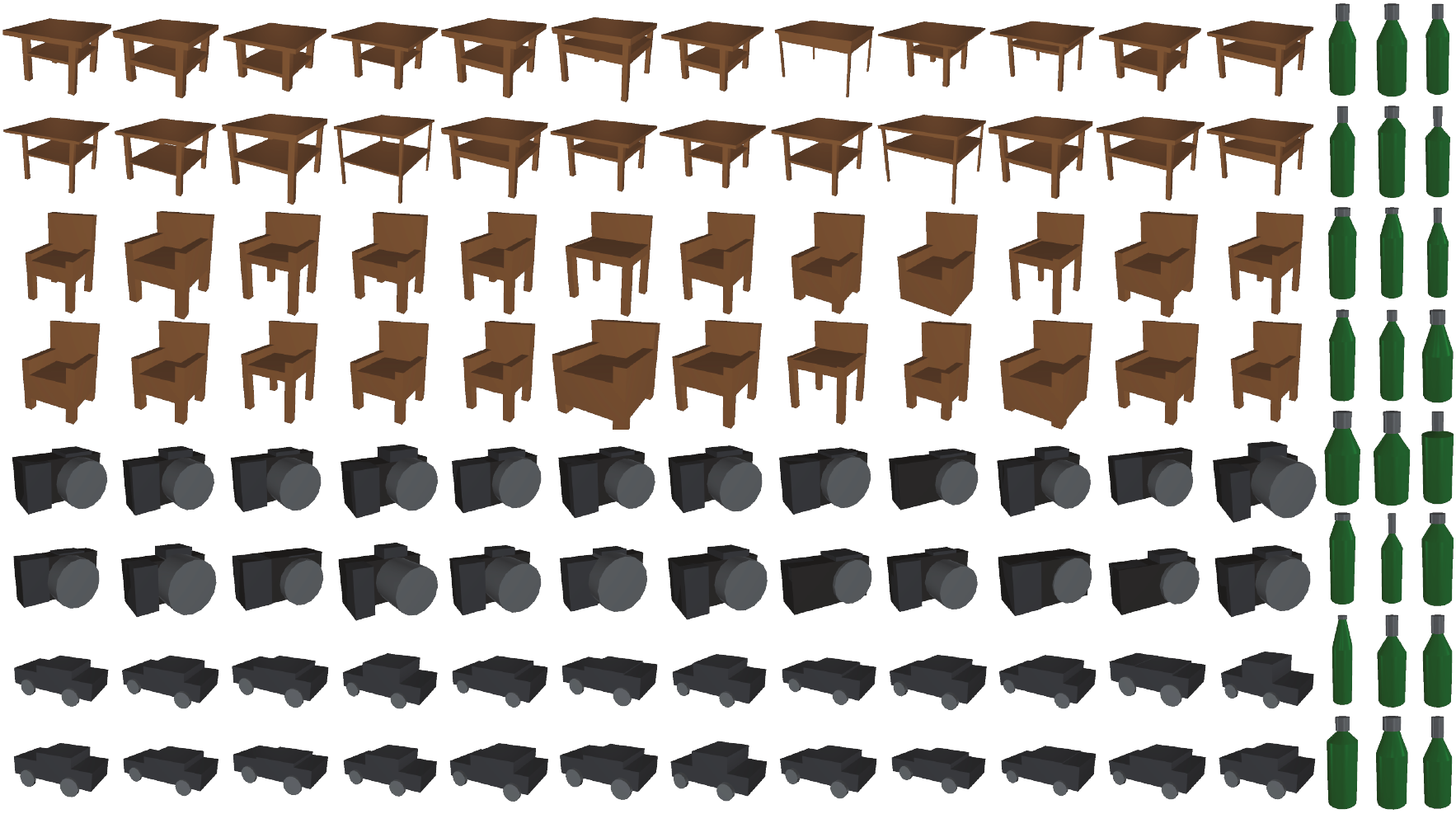}
    \caption{A gallery of design variations using the constrained inferred from our pipeline.}
    \label{fig:constraint-gallery}
\end{figure*}

We show the results of running our pipeline on five initial models: chair, table, car, camera, and bottle in Figure \ref{fig:constraint-gallery}. We handpicked five prompts from a list generated by ChatGPT and manually added a ``bench'' prompt for the chair example. For the differentiable renderer, we use Nvdiffrast~\cite{Laine2020diffrast}. We use stable diffusion model v1.4 from \cite{rombach2022highresolution}, and we use the mesh boolean algorithm from \cite{CPAL22} to compute the IoU. All experiments were conducted on a machine with an NVIDIA 2080Ti GPU. 

\begin{figure*}[ht!]
    \centering
    \includegraphics[width=\linewidth]{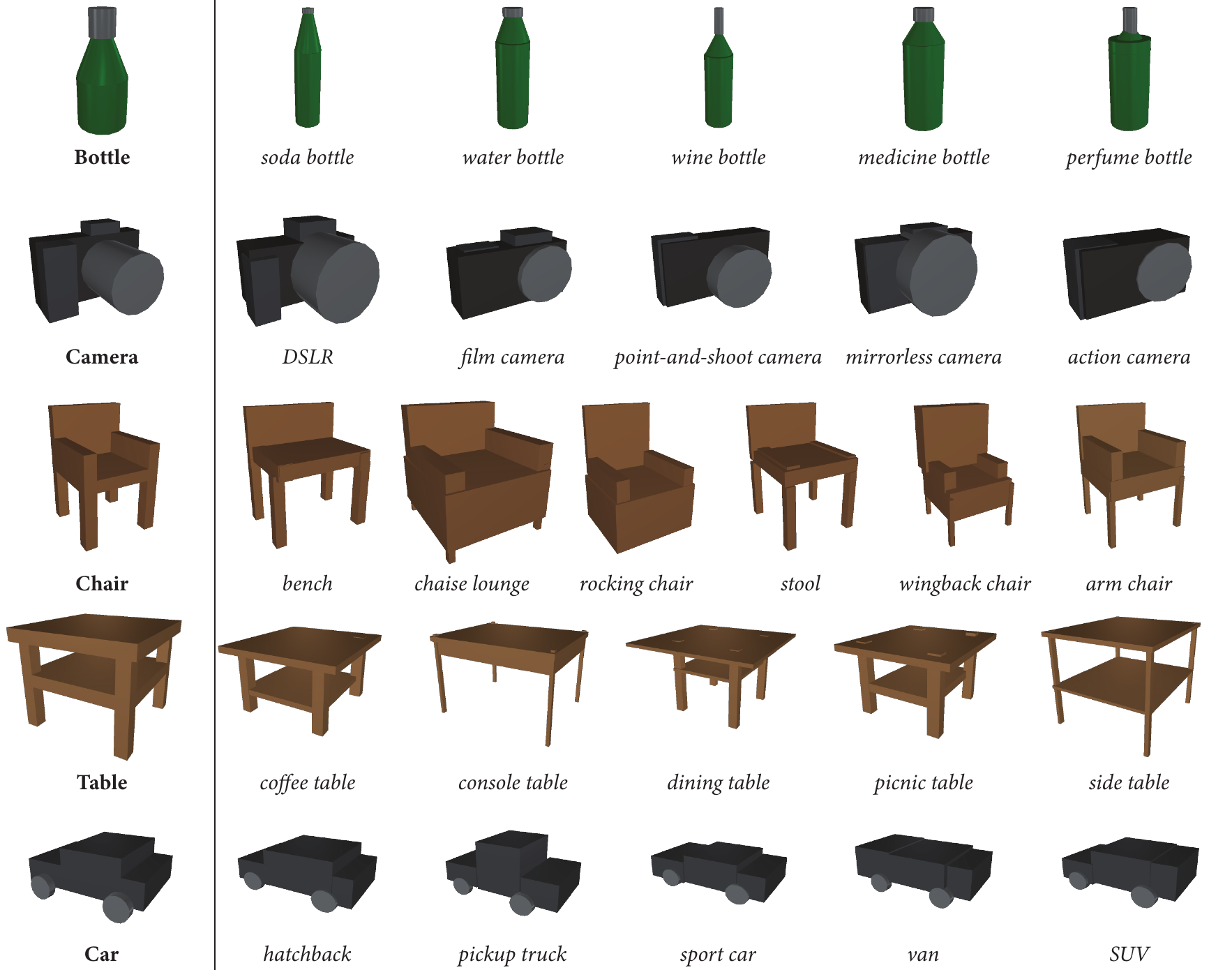}
    \caption{A gallery of text-condition CAD parameter optimization of different input models (first column) towards varying prompts.}
    \label{fig:generation-gallery}
\end{figure*}

\subsection{Generating CAD Variations}
In Figure \ref{fig:generation-gallery}, we demonstrate our method's ability to generate variations of the input model that are coherent with the text prompt with varying geometry and complexity. For example, we can observe that the bench is wider and has no arms while the chaise lounge has arms and is bulkier and that the SUV has a tall back section while the pickup truck has a low back section. 
We also notice that the method is able to discover part of the program that can be removed, for example, the arms of the chair for ``stool'' and the camera grip for ``action camera'' disappear by shrinking and blending into the rest of the shape. 

The method's inclination of staying close to the input model and within the parameter space leads to some intriguing results. For example, the ``rocking chair'' prompt produces a model resembling a nursing glider, which belongs to the same category and can be represented using the input CAD program. Similarly, the ``dining table'' generated by the method exhibits a surprising square shape and middle shelf. While these may not conform to the conventional idea of a dining table, such dinning tables exist in reality, making them feasible options which are chosen due to their proximity to the input model (see Fig.~\ref{fig:misc} (d))

Overall, our method performs well in maintaining the overall structure of the shapes while allowing unique changes that define each of the variations. 
We notice however, that there are some artifacts from these types of AI pipeline and gradient descent optimization. Notably, these generated models contain imperfections on primitive alignment (see corners of the chairs or table where the legs do not perfectly line up with the seat). These imperfections must be captured and cleaned up by our symbolic approach.

\subsection{Inferring Constraints and Reparameterization}

\begin{figure}
    \centering
    \includegraphics[width=\linewidth]{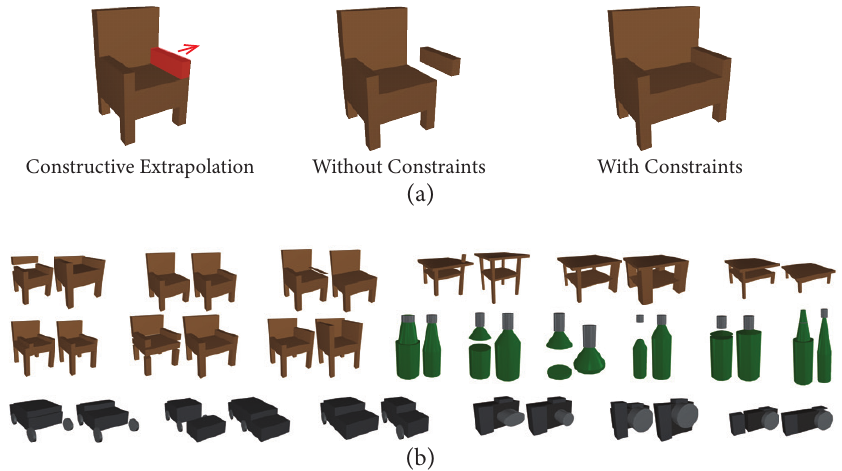}
    \caption{(a) Applying an extrapolative edit to a model can lead to broken geometry, but the constraints we discover prevent this and result in meaningful global edits from local changes. (b) A selection of random extrapolative edits without constraints (left) and with (right). Discovered constrains maintain geometric coherence. }
    \label{fig:goldilocks}
\end{figure}

\begin{figure*}[ht!]
    \centering
    \includegraphics[width=\linewidth]{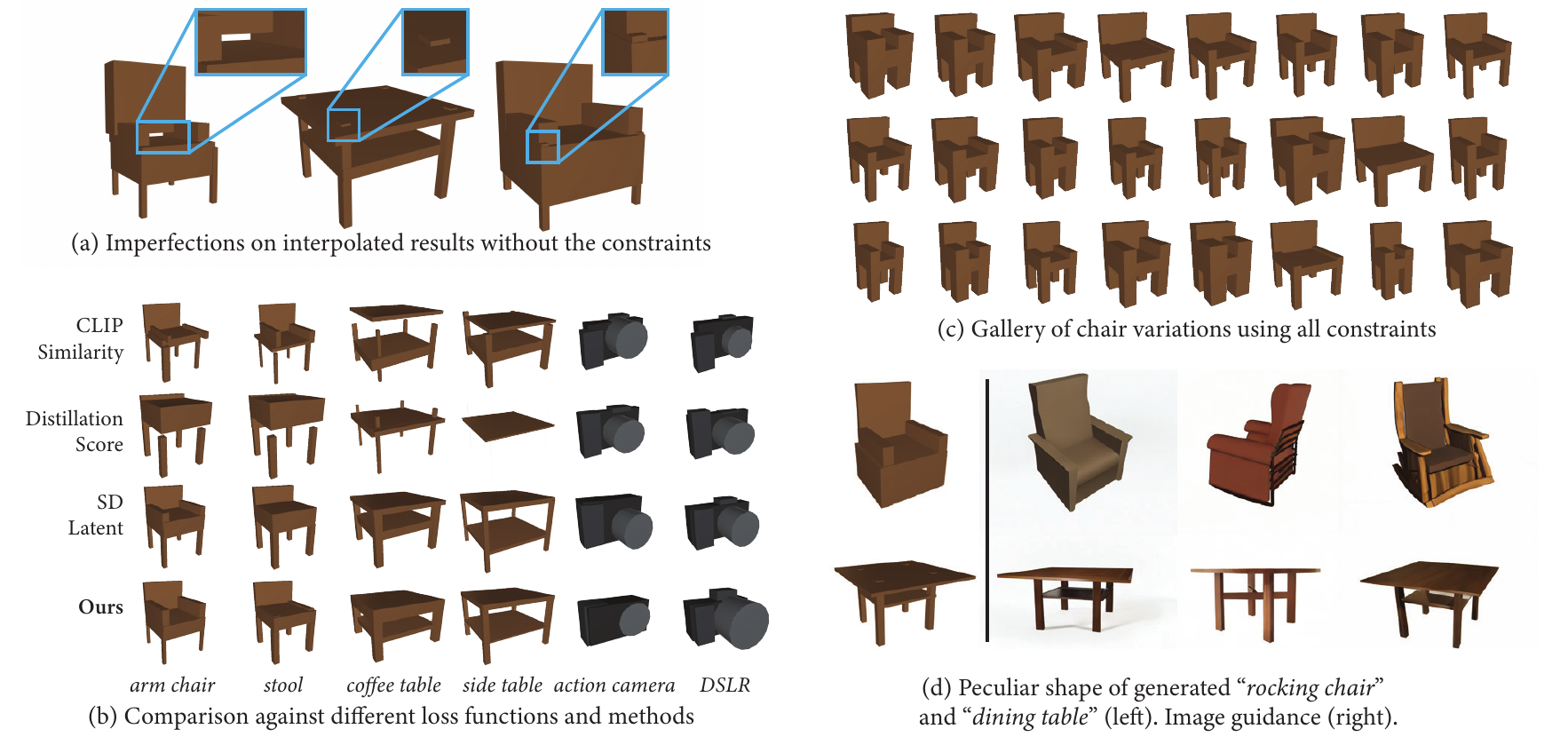}
    \caption{(a) Imperfections on interpolated results without the constraints. (b) Generation results using different loss functions: CLIP similarity, distillation score, L2 difference of the image latent of the rendered and diffusion-generated images, and L2 difference in the images (c) A gallery of chairs when all constraints of the base model are imposed. (d) Peculiar shape of generated ``rocking chair'' and ``dining table'' (left) and the image guidance (right).}
    \label{fig:misc}
\end{figure*}

Running our constraint discovery algorithm on the input models resulted in a dimensionality reduction from 19-28 parameters to 9-15, summarized in Table~\ref{tab:number-constraints}. We find semantically meaningful constraints; for example car wheels are co-axial, chair arms stay at the sides of the seat, and table and chair legs are all the same height. Our algorithm also rejects overly specific constraints such as the chair's seat being square and the same thickness as the arms and legs, which would overly constrain the design space as seen in Figure~\ref{fig:misc}~(c). The variety and quality of generated models on display in Figure~\ref{fig:constraint-gallery} shows that our constraints strike a balance between restriction and expression. They also have the effect of removing noise present in the input models, which propagates to the interpolated results as shown in Figure~\ref{fig:misc}~(a). On top of interpolation with manipulation parameters, we also support extrapolation by direct control of free construction variables. As Figure~\ref{fig:goldilocks} shows, extrapolation without constraints often produce incoherent geometry, but discovered constraints can prevent this.

\subsection{Ablations of our Neurosymbolic Approach}

Finally, we formalize the need for our neurosymbolic approach by comparing it to purely neural and purely symbolic techniques. A purely neural approach would only consider the generated model without any constraints. As seen in Figure~\ref{fig:misc}~(a), this would result in many artifacts.

On the other hand, if we apply all the geometric-semantic constraints inferred from the original model without using the generated images as input, we would end up with a more constrained space, as shown in Figure~\ref{fig:misc}~(c); most of the chairs are simply scaling variations that can appear overly boxy or skewed. 
We note that the variations we shown in Figure~\ref{fig:misc} (c) are still leveraging the bounds that we have discovered with our technique. In practice just imposing constraint does not define a bounded space for direct manipulation, and infinitely large boxy results can be generated---indeed these methods are used in companion with variational techniques for interaction~\cite{iWires:2009}.

\subsection{Ablations of our Text-Conditioned Variation Generation}\label{sec:ablation}

Figure~\ref{fig:misc}~(b) shows optimized CAD models using different loss functions: (1) CLIP similarity loss; (2) distillation scores, where the gradient is the difference between the added noise and predicted noise in latent space; (3) L2 difference in the image latent by the autoencoder; and (4) our image-space loss. For CLIP, we believe the unfavorable results is due to how the gradient is propagating back to the low-dimensionality of the CAD parameters. For stable diffusion, we observe that a full generation process is necessary to get an effective image guidance as oppose to using the difference between a single denoise step as seen with the distillation score.
Using the full diffusion process, whether using the image similarity or its latent, better preserves the overall structure and connectivity and converges to a desirable shape variation.
We decide to use the image difference to avoid having to backpropagate to the image encoder.

\begin{figure*}[ht!]
    \centering
    \includegraphics[width=\linewidth]{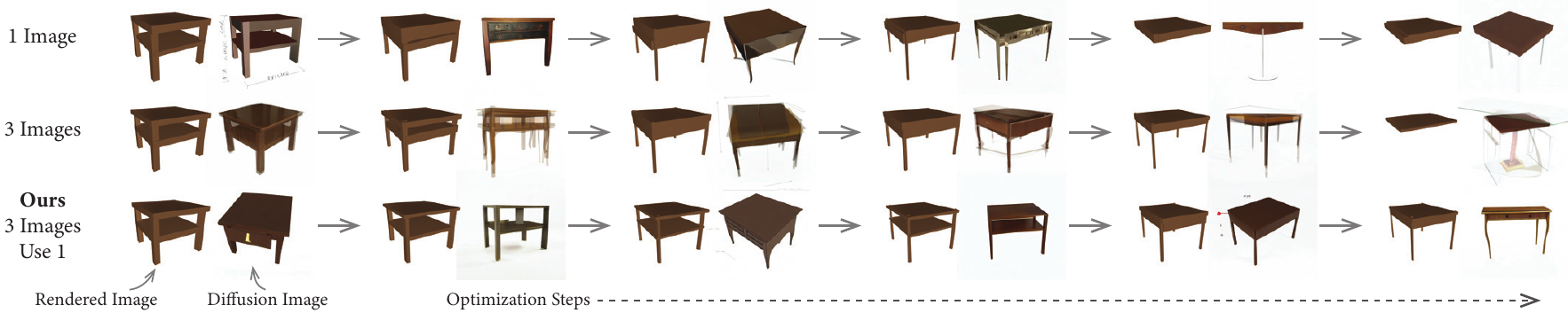}
    \caption{Rendered and diffusion-generated images at different iterations during optimization using (top) one diffusion-generated image per camera pose; (middle) three images; and (bottom) three images and select one image that is closest to the rendering. Note that the camera angles for the diffusion-generated images are random.}
    \label{fig:ablation-num-sd}
\end{figure*}

In Figure~\ref{fig:ablation-num-sd}, we show the optimization process of methods using different number of diffusion-generated images per camera pose. As discussed, a stable diffusion process will generate slightly different images at each execution. This causes thin parts like the legs more susceptible to this randomness; for example, if the legs in the generated images does not overlap at all with the rendering or partially overlap in different directions, the differentiable render will think the leg should not be there or should be thinner in all direction. We observe that using only the generate image that is the closest to the rendering reduce this effect and help maintain thin structures.

\begin{table}[h!]
\caption{Number of parameters, number of constraints in the base model, number of inferred constraints, number of reparameterized dimensions for each model}
\resizebox{\linewidth}{!}{%
\begin{tabular}{ccccc}
\toprule
Model & Parameters & \makecell{Base\\Constraints} & \makecell{Common\\Constraints} & \makecell{Constraint\\Dimensionality} \\
\midrule
Bottle & 19 & 20 & 8 & 9 \\
Camera & 24 & 21 & 9 & 15 \\
Chair & 48 & 172 & 37 & 11 \\
Table & 36 & 90 & 22 & 14 \\
Car & 42 & 94 & 26 & 13 \\
\bottomrule
\end{tabular}%
}
\label{tab:number-constraints}
\end{table}

\section{Limitations and Future Work}

\paragraph{Extending CAD Language Support} Our current implementation of our method is limited to a subset of the full CAD language. This limitation carries over from mesh differentiable renderer that we build upon. However, we believe that the techniques we propose can be extended to support missing CSG operations, such as difference and intersection, by either employing an SDF differentiable renderer or by integrating with differentiable CAD kernels.

\paragraph{Generating Complex 3D Geometries} The vanilla stable diffusion method encounters challenges in generating complex 3D geometries due to the lack of a strong 3D prior and limitations in view conditioning. Recent advances in the field have proposed promising strategies to bridge the gap between 2D and 3D representations \cite{xu2023dream3d, nichol2022pointe, seo2023let}. Incorporating these strategies could potentially enhance the generation of complex 3D geometries in our approach. Furthermore, it is worth noting that the computational cost associated with this step is considerable, and further efforts to reduce computational overhead would be valuable.

\paragraph{Handling Noise in Design Variations} The presence of artifacts in the image sampled from the diffusion model as well as the imperfect alignment of the model with the image can introduce errors into the generated variations. A fundamental challenge during constraint discovery lies in distinguishing these errors from intentional design variations. Incorporating additional considerations, such as physical understanding, can aid in tackling this issue. Furthermore, alternative approaches that utilize foundational models to differentiate between these types of variations can be envisioned. Finally, our algorithms for constraint discovery and reparametrization rely on the linearity of constraints. Extending our method to discover non-linear constraints would require addressing the additional computation cost associated with non-linearity.

\paragraph{Quality of Text Prompts} The quality of the results generated by our approach is heavily influenced by the quality of the available text prompts. When the requested shapes exceed the capabilities of the model, the performance may be subpar or lead to unexpected outcomes. Furthermore, we observe that prompts containing descriptive adjectives like "tall" or "wide" yielded minimal variations, while specifying object types proved to be more effective in generating diverse outcomes. ChatGPT may not always generate prompts that the user wants to explore. For instance, we had to manually add ``bench'' as a variation we would like to explore for the chair model. Future advancements in prompt engineering and involving users in the process can greatly improve the generation of meaningful variations and enable re-parameterization with higher semantic relevance. 

\paragraph{Fixed Program Structure} Our current implementation adheres to the structure found in the input CAD program, limiting design variations to those that can be realized solely through parameter tweaking. However, the input model may not necessarily represent the ideal base model for exploring the shape class. Recent advancements in program synthesis techniques have opened the door to complex automatic program transformations guided by semantics~\cite{liftinghalide,replsynthesis} or a handful of examples~\cite{flashfillpp}. Integrating these techniques into our system holds great promise, as it would enable us to not only modify parameters but also adjust the program structure itself to better align with user intent.
\section{Conclusion}
In this paper, we present a novel approach that leverages foundational models to facilitate CAD program manipulation. While prior applications of foundational models have focused on automatic 3D shape generation, our approach breaks new ground by utilizing them for structural shape representations. Our approach demonstrates the potential of integrating AI models with symbolic program analysis techniques and opens up exciting avenues for future research in the CAD design domain. 

{\small
\bibliographystyle{ieee_fullname}
\bibliography{bibliography,schulz,cad,textto3d}
}

\end{document}


\title{Supplemental: ReparamCAD: Zero-shot CAD Program Re-Parameterization for Interactive Manipulation}

\author{Milin Kodnongbua}
\email{milink@cs.washington.edu}
\affiliation{%
  \institution{University of Washington}
  \country{USA}
}
\author{Benjamin T. Jones}
\email{benjones@cs.washington.edu}
\affiliation{%
  \institution{University of Washington}
  \country{USA}
}
\author{Maaz Bin Safeer Ahmad}
\email{mahmad@adobe.com}
\affiliation{%
  \institution{Adobe Research}
  \country{USA}
}
\author{Vladimir G. Kim}
\email{vokim@adobe.com}
\affiliation{%
  \institution{Adobe Research}
  \country{USA}
}
\author{Adriana Schulz}
\email{adriana@cs.washington.com}
\affiliation{%
  \institution{University of Washington}
  \country{USA}
}

\renewcommand{\shortauthors}{Jones et al.}




\maketitle

\section{Implementation Details}
\subsection{Parameters for our CSG}
We implemented three primitives: cubes, cylinders (for each 3 axes), and cylinders with changing top radius. Each primitive has an associated translation $(t_x, t_y, t_z)$ and scale $(s_x, s_y, s_z)$ parameters. To convert a primitive to a mesh, we start from a base mesh of that primitive in a origin-centered, unit cube from which we apply the scale and translation transformation. For cylinder with changing top radius, we expose another parameter $r_\text{top}$ which controls the top radius of the base cylinder (before scaling and translation).

\subsection{Algorithms for Imposing Constraint}

As described in the main paper, we want to project parameters to a constrained space that will maximize the IoU between the constrained and unconstrained 3D shape. Since our simplified CSG programs allow overlapping primitives, we would require a differentiable mesh boolean algorithm which is an open research problems. We propose two algorithms. The first one makes a simplifying assumption that primitives cannot change in shape (a cylinder cannot change its top radius). It is based on linear transformation and is fast to compute. This assumption does not hold for the bottle which have a changing top radius due to it not being linear in the way we parameterize it, so we propose another algorithm that optimizes image-space loss via a differentiable renderer.

\subsubsection{Algorithm 1}

Given a CAD parameters $\mathbf{x_0} \in \mathbb{R}^{d}$ and a constraint matrix $C \in \mathbb{R}^{m \times d}$, where $m$ is the number of constraints and $d$ is the dimensionality of the input CAD model, we will describe a way to project $x$ to the constraint manifold that will approximately maximizes the IoU score.
Since we assume the base geometry of a primitive cannot change, we can use its canonical bounding cube as a proxy.

We observe that to maximize IoU, we need faces to stay at the same location as possible and we want to move faces that would cause minimal volume changes if necessary. 
We formulate the cost of shifting a face from its unconstrained position in its normal direction as the squares of the volume change generated by the shift. For cubes, the amount of volume change would be equal to the area of the face multiply by the amount of shifting. For cylinder, we additionally multiply that by the volume ratio (i.e. $\pi 0.5^2 = \pi / 4$).

Let $Q \in \mathbb{R}^{6P \times d}$, where $P$ is the number of primitives in the model, be a mapping from parameters to reduced coordinates of each face (a face on an $xy$-plane is represented by its $z$ coordinate, etc.); and let $\mathbf{a} \in \mathbb{R}^{6P}$ be the volume change associated with each face. We want to minimize $\lVert AQ\mathbf{x} - AQ\mathbf{x}_0 \rVert^2$ where $A := diag(\mathbf{a})$.

To minimize this on a constrained manifold, we first find the null space $N \in \mathbb{R}^{d \times d'}$ of the constraints matrix $C$ where $d'$ is the nullity of $C$. We can now formulate the optimization as:
\begin{align*}
    y^* := \argmin_y \lVert AQNy - AQx \rVert^2
\end{align*}
where $y \in \mathbb{R}^{d'}$ is the reduced coordinates.
Our projected parameters is $x^* := Ny^*$.

\subsubsection{Algorithm 2} We optimize the parameters based on the image difference loss. We first uniformly sample the camera poses and have that fixed for all subsequent runs. We optimize:
\begin{align*}
    y* := \argmin_y \sum_\alpha \lVert R_\alpha(Ny) - I_\alpha \rVert^2
\end{align*}
where $\alpha$ is a camera pose, $R_\alpha(\cdot)$ is the differentiable renderer, and $I_\alpha$ is the image we optimize towards. This can be images from diffusion or the initial renderings (i.e., $R_\alpha(\mathbf{x}_0)$).


\bibliographystyle{ACM-Reference-Format}
\bibliography{bibliography,schulz,cad}